\documentstyle[psfig]{l-aa}
\newcommand {\vdp} {{\em vdp~}}
\begin{document}
\thesaurus{ 11.03.4 Coma cluster; 04.03.1; 11.03.1; 11.04.1 }
\title{Segregations in clusters of galaxies
	\thanks{http://www.astrsp-mrs.fr/www/enacs.html}
}
\author{C.~Adami \inst{1}, A. ~Biviano \inst{2},
A.~Mazure \inst{1}}
\institute{IGRAP, Laboratoire d'Astronomie Spatiale, 
Traverse du Siphon, F-13012 Marseille, France  
\and Istituto TeSRE, CNR, via Gobetti 101, I-40129 Bologna, Italy}
\offprints{C.~Adami} 
\date{Received date; accepted date} 

\maketitle 
 
\markboth{Segregations in clusters of galaxies}{} 
 
\begin{abstract} 
We analyse a sample of about 2000 galaxies in 40 regular clusters, 
to look for evidence of segregation with respect to
galaxy luminosities and morphological types. 

We find evidence of luminosity segregation for galaxies brighter than 
$M_R < -21.5$, i.e. typically
the four brightest members of each cluster. We also find
evidence of morphological segregation: both the core-radius and the
velocity dispersion increase along the Hubble sequence (ellipticals - S0 - 
early spirals - late spirals). 

Galaxies of different types have different velocity dispersion profiles,
being steeper for later type galaxies. Simple modelling allows us to show
that elliptical (and, to a lesser extent, S0) orbits are mostly tangential 
in the cluster core, and nearly isotropic outside, while spiral
(in particular late-spiral) orbits are predominantly radial.

A viable interpretation of our results is that (1) late spirals,
at variance with other type galaxies, are a non-virialized cluster population, 
still on partially radial infalling orbits, (2) the elliptical phase-space
distribution is evolving towards energy equipartition through the process of
dynamical friction, (3) S0 and early-spirals have intermediate distributions
between these two extremes.
\end{abstract} 
 
 

\section{Introduction} 
The analysis of the phase-space distributions of cluster galaxies helps
to constrain models of galaxy formation and evolution. In particular,
it is important to distinguish among galaxies of different luminosities and
morphological types, as they may not have formed and evolved in the same way.
Moreover, it is important to know the phase-space distributions of
different cluster galaxies for a proper determination of the cluster mass
(Girardi et al. 1996).

The observational evidence that early-type galaxies occupy denser 
environments than late-types dates back to Curtis (1918) and Hubble \& Humason
(1931), and was quantified by Oemler (1974). Melnick \& Sargent (1977) 
showed that the relative fraction of S0 and spiral galaxies depends
on the distance from the cluster centre. Most subsequent studies were based
on the data sample of Dressler (1980a) containing over 6000 galaxies in 55
cluster fields. While Dressler (1980b) concluded that the basic correlation
is between morphology and {\em local density} in clusters, Sanroma \& 
Salvador-Sole (1990) and Whitmore \& Gilmore (1991), based on the same
sample, have independently reached the conclusion that the fundamental
correlation is between morphology and {\em global cluster properties,} such as
the the clustercentric distance. For more complete reference, we refer
the reader to Whitmore et al. (1993).

More recently, ellipticals and S0 were found to have smaller velocity
dispersions ($\sigma$ in the following)
than spirals and irregulars (Tammann 1972; Melnick \& Sargent 1977;
Moss \& Dickens 1977; Sodr\'e et al. 1989; Biviano et al. 1992, hereafter B92; 
Stein 1997, hereafter S97). As morphologies are usually
more difficult to reliably determine than colours, and colours and morphologies
are correlated, many authors have analysed the correlation between kinematics
and colours. Andreon (1996), Biviano et al. (1996),  Colless \& Dunn (1996),
all found that the ratio between the $\sigma$ of the blue and the red galaxy
populations in the Coma cluster is $\sim 1.3$--1.4. Similar evidence has
recently been found by Carlberg et al. (1997) in their analysis of the CNOC
survey medium-z clusters. Based on a sample of $\sim 600$ galaxies 
in 15 clusters, S97 has recently confirmed the larger $\sigma$ for
spirals as compared to early-type galaxies, and has shown that within the class
of early-type galaxies, S0 have a larger $\sigma$ than ellipticals.
 
Further, possibly related, observational evidence is the existence of
luminosity segregation: the most luminous galaxies are closer to 
the cluster centres (e.g. Rood \& Turnrose 1968, Capelato et al. 1981, 
Yepes et al. 1991), and have a significantly lower velocity dispersion 
(Rood et al. 1972, Chincarini \& Rood 1977, Struble 1979, Kent \& Gunn 1982, 
B92, S97). The anti-correlation between luminosity 
and velocity dispersion seems however to hold for the very bright galaxies only
(luminosity $\geq 5 L^{\star}$), and to depend on the galaxy morphological 
type, being strongest for ellipticals (B92, S97).
S97 in particular has pointed out that there is no evidence whatsoever
for a $\sigma$--luminosity (anti-)correlation in galaxies other than
ellipticals.

Only very recently, large data-sets have allowed the analysis of the velocity
dispersion profiles (\vdp in the following) of galaxies of different
properties. Mohr et al. (1996) have shown that the emission-line galaxies in 
the A~576 cluster have a steeper \vdp than non-emission-line galaxies; 
Biviano et al. (1997, hereafter B97) have come to the same conclusion on the 
basis of the ENACS data-set of 75 clusters (see Katgert et al. 1996, 
hereafter K96). A similar difference is found between the \vdp of red and blue 
galaxies in 15 clusters from the CNOC survey (Carlberg et al. 1997).

From the theoretical side, the evidence for segregation is interpreted either
as arising in 
a different formation process for galaxies of different luminosities and/or
morphologies, or as 
being due to a different evolutionary history (the "nature vs. nurture"
problem). Many physical processes can affect the morphology, luminosity,
and velocity of a galaxy. Dynamical friction can slow down the more massive
galaxies, circularize their orbits (den Hartog \& Katgert 1996 and
references therein), and enhance their merger rate (Mamon 1992, 1996 and
references therein); ram pressure and collisions can transform a star forming 
galaxy into a quiescent one; tidal effects can
truncate the galaxy sizes and reduce their luminosities while producing a
reservoir of debris for centrally located cD galaxies (Oemler 1992, 
Whitmore 1994, and references therein). The steep \vdp and
large $\sigma$ of star-forming galaxies, as compared to quiescent ones,
suggests that star-forming galaxies are falling into the cluster, maybe
for the first time (B97).

In this paper we revisit these issues, via the analysis of a large data-set
of about 2000 cluster galaxies, partly based on the ENACS catalogue (K96,
Katgert et al. 1997), and partly drawn from the literature. 
Taking a step further
from previous studies on segregation, we distinguish spirals into two
classes (early- and late-type), and determine the \vdp of 
ellipticals, S0, early-spirals and late-spirals.

The structure 
of the paper is as follows: in \S~2 we describe the data-sample, in 
\S~3 we address the issues of segregation in velocities and spatial segregation,
and determine the \vdp for different morphological classes;
in \S~4 we discuss our findings, and in \S~5 we give our conclusions.

\section{The data}
\subsection{The compilation of the catalogue}
The data sample on cluster galaxies which we use in this paper
is a merging of the ENACS data-base (K96 and Katgert et al. 1997)
and available data from the literature, which includes the December 
1994 version of ZCAT (Huchra et al. 1995). Coordinates and redshifts are
available for all galaxies in our data sample, and magnitudes and morphological
types are available for most of them.

In compiling this data-base, we had to make a compromise between 
collecting as much data as possible, and keeping the data-base 
homogeneous. For this reason, 
before putting together data from different sources, we 
first compared the different redshift estimates for galaxies common to 
different data-sets. Only when the average velocity
difference of all common galaxies was not statistically 
significant (at the 90~\%
confidence level) and less than 200~km~s$^{-1}$,
were the different source redshift scales considered compatible and the 
data-sets merged together. For a comparison of ENACS and literature data
we refer the reader to K96.

In merging different data-sets, one has
to take special care not to insert the same galaxy
twice. This may happen if
the galaxy coordinates in either of the two data-sets are not accurate enough.
For the ENACS clusters, we made
use of the (unpublished) catalogue of galaxies obtained by scanning film copies
of the SERC blue survey and glass copies of the first Palomar Sky Survey,
with the {\em Astroscan} measuring machine (K96 and references
therein). This catalogue was used in preparing the Optopus observations; for all
detected galaxies a photometric parameter was derived, which was later
converted into an isophotal magnitude ($R_{25}$) using CCD-observations for
calibration (more details are given in K96). This photometric catalogue
is meant to be complete, but only for selected cluster regions specifically
chosen for the Optopus observations. When a galaxy with a redshift measurement
from the literature was located in the Optopus plate cluster regions, 
it was possible to find a corresponding entry in the {\em Astroscan} catalogue.
If the galaxy already had an ENACS redshift measurement, this was adopted;
otherwise, the literature redshift was used, but the coordinates and 
magnitudes were taken from the {\em Astroscan} catalogue, for a better 
consistency with other ENACS galaxies. 

For the non-ENACS clusters, or for regions of the ENACS cluster not covered by
the Optopus plates, we took a conservative approach and
rejected all suspected double entries for the same galaxy. The decision was
based upon similarity of coordinates (up to 30" difference),
and galaxy properties like redshift, and/or magnitude, and/or morphology. 
We may have eliminated galaxy pairs,
but this is not a concern for the present analysis, as long as 
the probability for galaxies to be in pairs does not depend strongly on their
morphological type and/or luminosities.

The average galaxy velocity uncertainty in our final data-set,
using the errors quoted in the original references, is
69~km~s$^{-1}$, and the maximum quoted uncertainty is 276~km~s$^{-1}$.
External errors are likely to be larger, due to
the non-perfect match of the redshift scales of merged data-sets. 

Magnitudes were corrected for galactic absorption, using the
extinction-maps of Burstein \& Heiles (1982), and the relations between
$E(B-V)$ and absorption in several magnitude bands as given by
Colless (1989), Oemler (1974), and Postman \& Lauer (1995). We applied 
k-corrections using the relations given by Colless (1989), Postman \& Lauer 
(1995), Sandage (1973), and Schneider et al. (1983). All magnitudes were then
converted to ENACS $R_{25}$ magnitudes.
When possible, the linear regression 
between the ENACS and the literature magnitudes for common galaxies was used as
a conversion formula. Otherwise, we used relations given in the literature to 
convert from total to isophotal magnitudes, when needed, and 
from different bands to the red bandpass
(ref.s mentioned above and:  Dickens et al. 1986, Lugger 1989,
Oegerle \& Hoessel 1989, and Sharples et al. 1988). 

The uncertainty of ENACS magnitudes is $\leq 0.3$~mag (and even lower for 
galaxies in clusters with CCD-calibration, see K96). The uncertainty 
of the magnitudes taken from the literature is probably larger ($\leq 1$~mag),
because of the additional error introduced in transforming them to the ENACS
bandpass.

All the morphological type estimates were taken from the literature, 
as there is no morphological information available in the ENACS data-set.

\subsection{The definition of clusters}
Several methods have been proposed in the literature for defining cluster
membership. As a large part of our data comes from ENACS, it would be
natural to follow K96's methodology of fixed gaps: redshift boundaries for
a cluster are defined such that the closest non-cluster member in redshift
space is separated by at least 1000~km~s$^{-1}$ from either of the two 
boundaries. However, a fixed gap of 1000~km~s$^{-1}$ is a valid choice only
for the relatively homogeneous ENACS sample, but cannot work for all
data-sets, simply because the average density of observed galaxies per redshift
interval may change (as a consequence of one survey being deeper than another).
As an example, the fixed gap fails to effectively identify the Coma
cluster in redshift space.

For this reason, we adopted a gap that depends on the galaxy density per
redshift interval along the line-of-sight to a given cluster, and we call it
the density-gap, equal to $500+600 \exp (-n/33)$~km~s$^{-1}$, where $n$ is the
total number of galaxies with measured redshift
in the cluster region. This expression was derived 
from simulations of the occurrence of gaps of a given size in 
gaussian distributions with varying numbers of objects.
When applied on the ENACS data-base it identifies similar systems 
as those identified using the fixed gap criterion
(see also Adami et al. 1997a).

Using the density-gap method we identified the main systems along the lines
of sight. For the cluster A~548 alone we also applied a splitting in 
coordinate space, since it is clearly bimodal in projection (Escalera et al. 
1994).

In order to further reject possible remaining interlopers, we then applied the
method of den~Hartog \& Katgert (1996) to the systems with at least 50
galaxies (the method is unreliable for poorer data-sets).

\begin{table*} 
\caption[]{Parameters for the 40 clusters. The chosen centres are denoted by
"X" when they are X-ray determined, "cD" when they coincide with the position
of a cD galaxy, "I" when they are chosen at the peak of isodensity contours.
We show the number of bright galaxies for each cluster (see text). The 
references are:
A: Adams et al. (1980);
B1: Beers \& Bothun (1992);
B2: Beers et al. (1991);
B3: Biviano (1986);
B4: Bothun \& Schombert (1988); 
B5: Butcher \& Oemler (1985);
B6: Biviano et al. (1996) and references therein;
C1: Chapman et al. (1987);
C2: Chapman et al. (1988);
C3: Chincarini et al. (1981);
C4: Colless \& Hewett (1987);
C5: Colless (1989);
D1: Dickens et al. (1986); 
D2: Dickens \& Moss (1976);
D3: Dressler \& Shectman (1988);
D4: Dressler (1980a);
E: Ettori et al. (1995);
F1: Faber \& Dressler (1977);
F2: Fabricant et al. (1993);
F3: Fabricant et al. (1989);
F4: Fanti et al. (1982);
G1: Gavazzi (1987);
G2: Geller et al. (1984);
G3: Giovanelli et al. (1982);
G4: Gregory et al. (1981);
G5: Gregory \& Thompson (1978);
H1: Haynes (1980);
H2: Hill \& Oegerle (1993);
H3: Hintzen et al. (1982);
H4: Hintzen (1980);
K: Kent \& Sargent (1983);
L1: Lauberts \& Valentijn (1989);
L2: Lucey \& Carter (1988);
L3: Lucey et al. (1983);
M1: Malumuth et al. (1992);
M2: Moss \& Dickens (1977);
N: Nilson (1973)
O: Ostriker et al. (1988);
P1: Pinkney et al. (1993);
P2: Poulain et al. (1992);
P3: Proust et al. (1992);
Q1: Quintana \& Ramirez (1990);
R1: Richter (1989);
R2: Richter (1987);
S1: Scodeggio et al. (1995);
S2: Sharples et al. (1988);
T1: Tarenghi et al. (1979);
T2: Teague et al. (1990);
T3: Tifft (1978);
W: Willmer et al. (1991);
Z1: Zabludoff et al. (1993);
Z2: Zabludoff et al. (1990);
Z3: Zwicky et al. (1968)
} 
\begin{flushleft} 
\small 
\begin{tabular}{cccccccc} 
\hline 
\noalign{\smallskip} 
Name & Bright galaxies & Nb of redshifts & Mean redshift &   Extension    &  
$\sigma$ & Center & References \\ 
     &                 &                 &               & (h$^{-1}$~kpc) &
(km/s)   &        &            \\
\hline 
\noalign{\smallskip} 
A0119 & 1 & 73 & 0.042 & 994 & 943$\pm 170 $ & I & F2+ENACS \\ 
A0151 & 1 & 31 & 0.052 & 897 & 804$\pm 181 $ & I & D4+P3+ENACS \\ 
A0168 & 1 & 59 & 0.044 & 977 & 528$\pm 80 $ & I & F1+Z2+ENACS \\ 
A0194 & 0 & 52 & 0.017 & 965 & 523$\pm 208 $ & X & C2+B3 \\ 
A0262 & 0 & 40 & 0.016 & 996 & 506$\pm 110 $ & X & G3+G4+M2+Z3 \\ 
A0400 & 0 & 69 & 0.023 & 909 & 629$\pm 100 $ & X & B1+B5+D4 \\ 
A0426 & 0 & 75 & 0.017 & 798 & 1329$\pm 205 $ & X & B3+K+P2+Z3 \\ 
A0496 & 1 & 56 & 0.033 & 898 & 746$\pm 153 $ & X & D4+M1+Q1 \\ 
A0539 & 0 & 71 & 0.029 & 996 & 1095$\pm 390 $ & X & O+N+Z3 \\ 
A0548 & 0 & 88 & 0.041 & 992 & 736$\pm 84 $ & X & D3+ENACS \\ 
A0548 & 0 & 95 & 0.042 & 990 & 885$\pm 79 $ & X & D3+ENACS \\ 
A0576 & 0 & 26 & 0.038 & 670 & 927$\pm 244 $ & X & H1+H3+F4 \\ 
A0754 & 1 & 51 & 0.056 & 878 & 914$\pm 186 $ & X & D3+ENACS \\ 
A0978 & 1 & 18 & 0.056 & 992 & 674$\pm 181 $ & I & D4+ENACS \\ 
A0999 & 1 & 23 & 0.033 & 700 & 393$\pm 185 $ & cD & A+C1 \\ 
A1016 & 1 & 22 & 0.033 & 575 & 257$\pm 80$ & cD & A+C1 \\ 
A1060 & 2 & 112 & 0.013 & 951 & 665$\pm 78$ & X & R1+R2 \\ 
A1142 & 0 & 32 & 0.037 & 958 & 441$\pm 141 $ & X & B3+D4+G2 \\ 
A1367 & 0 & 50 & 0.023 & 863 & 878$\pm 255 $ & X & D2+G1+G5+T3+Z3 \\ 
A1631 & 1 & 43 & 0.048 & 996 & 706$\pm 159 $ & X & D3 \\ 
A1644 & 1 & 58 & 0.049 & 964 & 966$\pm 201 $ & X & D3 \\ 
A1656 & 1 & 196 & 0.024 & 974 & 1135$\pm 127 $ & X & B6 \\ 
A1983 & 2 & 42 & 0.045 & 903 & 593$\pm 101 $ & X & D3 \\ 
A2040 & 0 & 25 & 0.046 & 796 & 730$\pm 248 $ & X & Z2+ENACS \\ 
A2063 & 1 & 38 & 0.035 & 928 & 744$\pm 199 $ & X & B2+D4+H2 \\ 
A2147 & 0 & 22 & 0.037 & 971 & 1194$\pm 401 $ & X & T1+Z3 \\ 
A2151 & 0 & 74 & 0.037 & 983 & 861$\pm 122 $ & X & D3 \\ 
A2256 & 1 & 56 & 0.059 & 890 & 1403$\pm 221 $ & X & F3 \\ 
A2634 & 0 & 22 & 0.030 & 832 & 938$\pm 530 $ & X & B4+B5+D4+H4+ 
P1+S1+Z1 \\ 
A2670 & 2 & 95 & 0.075 & 978 & 1120$\pm 191 $ & X & S2 \\ 
A2877 & 1 & 37 & 0.023 & 865 & 790$\pm 376 $ & X & M1 \\ 
A3128 & 1 & 80 & 0.059 & 997 & 1041$\pm 162 $ & I & C4+C5+ENACS \\ 
A3158 & 5 & 33 & 0.059 & 778 & 1064$\pm 294 $ & I & C3+L2+ENACS \\ 
A3376 & 1 & 53 & 0.047 & 995 & 790$\pm 163 $ & X & D3 \\ 
A3381 & 1 & 26 & 0.038 & 777 & 306$\pm 145 $ & X & D3 \\ 
A3389 & 2 & 37 & 0.027 & 585 & 620$\pm 109 $ & X & D3+T2 \\ 
A3526 & 0 & 107 & 0.013 & 994 & 962$\pm 93$ & X & D1+L1 \\ 
A3574 & 0 & 33 & 0.016 & 881 & 513$\pm 115$ & X & W \\ 
A3716 & 0 & 57 & 0.044 & 966 & 845$\pm 117 $ & X & C4+D3 \\ 
A4038 & 1 & 42 & 0.028 & 771 & 872$\pm 288$ & X & E+L3 \\ 
\noalign{\smallskip} 
\hline	    
\normalsize 
\end{tabular} 
\end{flushleft} 
\label{t-data1} 
\end{table*} 
 
It is important to use only the central regions of clusters in our analysis, 
in order to minimize spatial selection effects. In denser regions, multi-object
spectroscopy -- from which most of our data comes -- tends to select brighter
galaxies than in less dense region. On the other hand, deeper surveys are done 
with smaller fields of view, which means that the central 
cluster regions are generally sampled to fainter magnitudes than 
the wider external regions. 
While a detailed analysis has shown that these selection effects are of no
concern for the ENACS data-set (Adami et al. 1997b), we do need to worry
about this issue when using literature data. To reduce this problem,
we chose to consider only galaxies within 1 h$^{-1}$~Mpc 
radius\footnote{Throughout the paper h$=$H$_0/(100 \, km \, s^{-1} \, 
Mpc^{-1})$, and the deceleration parameter $q_0=0$} from the 
cluster centres. Via this selection we also implicitly exclude regions 
that have not yet reached virialization (e.g. White 1992), yet which inlude
most of the cluster galaxies, as their projected density 
at 1~h$^{-1}$~Mpc is low, being $< 8$~\% of the central density for 80~\% 
of the clusters (Girardi et al. 1995). 
Centres were defined by choosing the peak of the X-ray emission, when
available, or the location of the cD, when present, or else the isodensity
contour peak (as in Adami et al. 1997b). 

As was shown by Beers et al. (1990), $\sim 20$ data-points
are sufficient to provide 
robust estimates of the central location and scale of a data-set, when the
biweight estimators are used (see also Lax 1985). On the other hand, Girardi et
al. (1993) have shown that changing the cluster membership criterion
changes the cluster $\sigma$ estimate if only 10 galaxies per cluster are
available. As a consequence, we considered in our analysis only 
those clusters with at least $\simeq 20$ galaxy members within a circle
of 1~h$^{-1}$~Mpc radius (the minimum number of cluster galaxies in our
sample is 18, in the cluster A~978).

The final sample contains 1997 galaxy members from 40 nearby clusters
(counting the subclusters in A~548 as two independent systems). The mean
cluster redshifts range from 0.013 to 0.075; the absolute R$_{25}$ magnitudes
of cluster galaxies range from -23 to -16; the morphological fractions of 
the 1997 galaxies are: 24~\% ellipticals (E, hereafter), 49~\% S0, 
21~\% early spirals (Sa to Sb in the Hubble sequence, globally labelled 
Se, hereafter), and only 6~\% late spirals or irregulars (Sl, 
hereafter). These fractions are similar to the average fractions for
all kinds of clusters (spiral poor, spiral rich and cD) as given by
Oemler (1974).

\begin{figure*} 
\vbox 
{\psfig{file=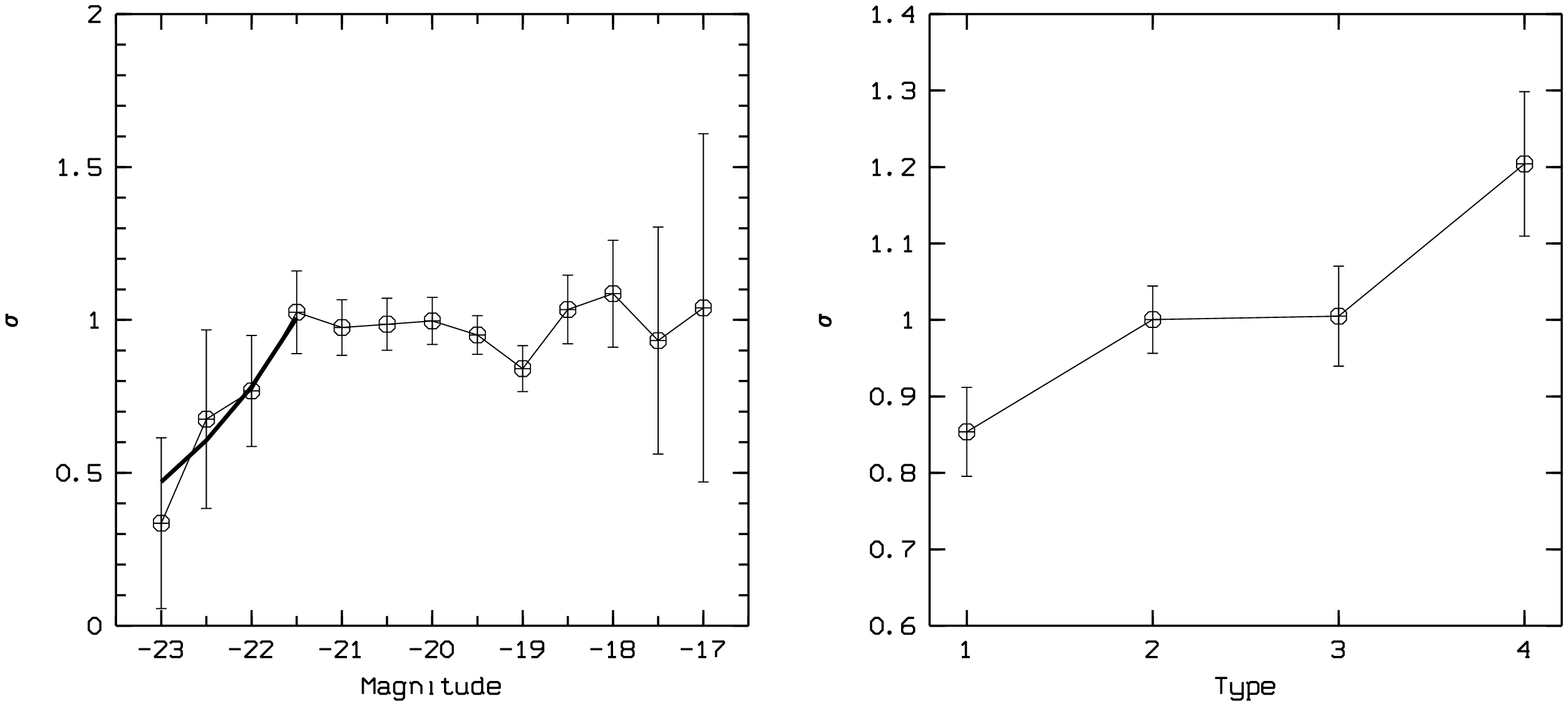,width=16.0cm,angle=270}} 
\caption[]{Left panel (a): normalized $\sigma$ vs. $M_R$ magnitude for all
galaxies in the synthetic cluster (1997 galaxies). We show the relation
$\sigma \propto 10^{0.2 M_R}$, normalized at $M_R = -21.5$; 
right panel (b): normalized $\sigma$ vs. morphological type for 
galaxies in the synthetic cluster with 1: E, 2: SO, 3: Se and 4: Sl.} 
\label{} 
\end{figure*} 

\begin{figure*} 
\vbox 
{\psfig{file=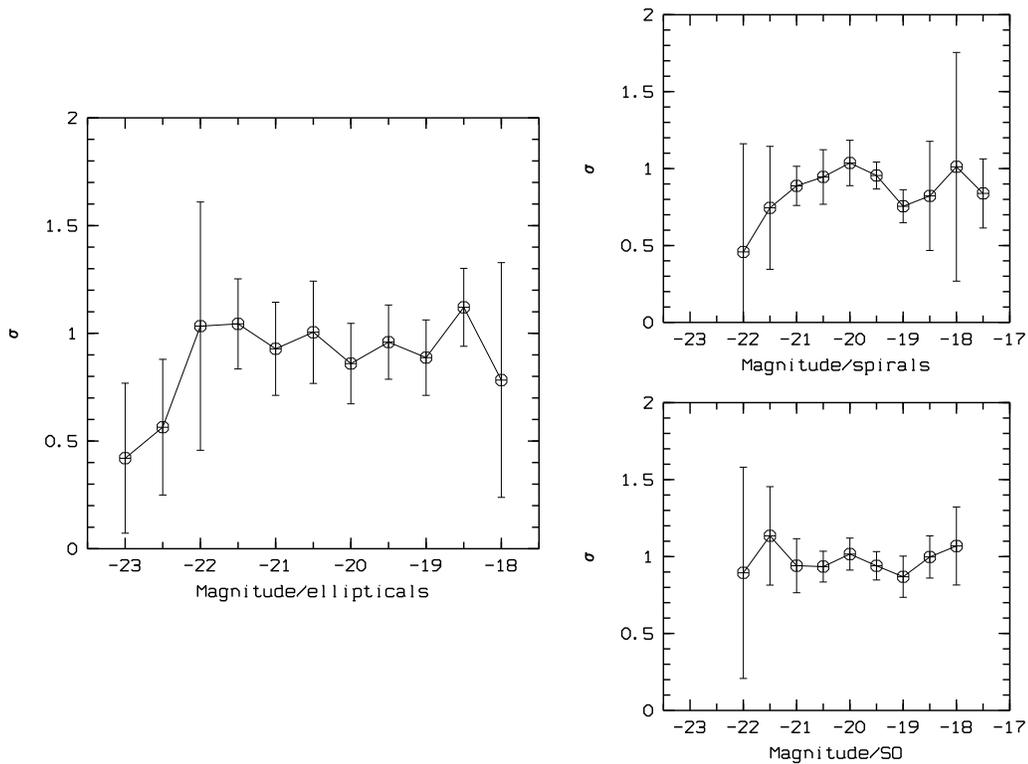,width=16.0cm,angle=270}} 
\caption[]{Normalized $\sigma$ vs $M_R$ magnitude for the galaxies
in the synthetic clusters, split into three morphological classes;
 left panel (a): E (479 galaxies); bottom-right panel (b): S0 (982 galaxies); 
top-right panel (c): spirals (536 galaxies).}
\label{} 
\end{figure*} 

In Table~1 we list for each system the cluster Abell
number, the number of very bright galaxies (i.e. those with a cD morphology
and those separated by a gap of at least 0.5 mag from the fainter galaxies),
the total number of redshifts, the mean redshift of the cluster, the
clustercentric distance that defines the region analysed in this paper,
the velocity dispersion and its bootstrap error based on
1000 resamplings, the type of
the selected centre ("I" means the centre was chosen at the isodensity peak,
"cD" that it coincides with the position of the cD galaxy, "X" that it
coincides with the position of the X-ray peak), and the relevant references
to the collected data.

While this paper was being written, the data-set of Stein (1996)
on cluster galaxies became available. As this data-set is not included
in our compilation, it makes sense to compare Stein's determinations
of the cluster parameters with ours.
There are five clusters in common with Stein (1996) in our data-set: 
A~400, A1016, A1060 (Hydra), A3526 (Centaurus) and A4038. As Stein's sample
is limited to the inner 0.5~h$^{-1}$~Mpc, we restricted our samples to this
inner region as well, for the sake of comparison. 
We find an average difference of 37~km/s in the $\sigma$ estimates, 
and of -122~km/s in the average velocity estimates, both 
within their 1-sigma uncertainties.
 

\section{Analysis and results}
Our results are based on the analysis of a synthetic cluster that
we have created by joining all 40 cluster samples (as in B92;
see also B97 and S97). In the synthetic cluster, galaxy distances are in
unit of h$^{-1}$~Mpc from the centre of the cluster they belong to.
Galaxy absolute magnitudes, $M_R$, are obtained from apparent magnitudes
by using the mean cluster distances, and galaxy
velocities are referred to the cluster average velocities and 
scaled by the cluster velocity dispersions,
\begin{equation}
v_{i}^{(n)}=\frac{v_i-<v>_j}{\sigma_j} 
\end{equation}
where $i$ and $j$ are, respectively, the galaxy and the cluster indices.

In order to check the results obtained on the synthetic cluster,
the same analysis was also performed on a single cluster, Coma (A~1656), 
for which a large enough data-sample was available.
The morphological fractions of Coma galaxies are
similar to those of the synthetic cluster.

In all following analyses, error-bars are obtained from a bootstrap
technique with 1000 resamplings.
 
\subsection{Segregation in velocity}

In order to test for the presence of luminosity segregation in velocity
space, we have computed the (normalised) velocity dispersion,
$\sigma^{(n)}$, of the synthetic cluster for galaxies of different 
(absolute) magnitude, without making any distinction according to the
morphological type. In Fig.~1a we plot $\sigma^{(n)}$ vs. mag using
0.5-mag bins; it can be seen that 
the velocity dispersion is roughly independent of magnitude for $M_R > -21.5$
but monotonically decreases at brighter magnitudes, from
$\sigma^{(n)} \simeq 1.0 \pm 0.1$ at $M_R=-21.5$, to
$\sigma^{(n)} \simeq 0.3 \pm 0.3$ at $M_R=-23$. The difference between
the velocity dispersions for galaxies brighter and, respectively, fainter
than $M_R=-21.5$ is significant at the 0.999 confidence level, 
according to an F-test. 

Galaxies brigther than this magnitude limit are
moving more slowly than other cluster galaxies. 
While it is well known that most cD and D galaxies
sit at the bottom of their cluster potential wells, there are at most 31 
such galaxies in our sample (see Table~1),
as compared to 166 galaxies brighter than 
$M_R=-21.5$; this effect thus concerns the 4 brightest galaxies
of each cluster on average.
This segregation in the velocity space may be interpreted as evidence
that the brightest cluster galaxies 
have reached energy equipartition as a 
consequence of dynamical friction (e.g. Capelato et al. 1981). 
For a constant galaxy mass-to-light ratio, energy equipartition implies
$\sigma \propto 10^{0.2 M_R}$. For $M_R \leq -21.5$ this relation is 
consistent with our data: the regression line between 
$\log \sigma ^{(n)} $ and $M_R$, has a slope of $0.22 \pm 0.06$.
The same relation (normalized at $M_R=-21.5$) is 
also shown superposed on the data points in Fig.~1a.
 
In order to analyse the dependence of luminosity 
on the galaxy morphological class, we have repeated the previous analysis
on the three subsamples of E, S0 and Se$+$Sl (early- and late-spirals and 
irregulars are considered together to avoid poor statistics).
The $\sigma$-magnitude relation for E is similar to that for all 
galaxies together (see Fig.~2a), and the $\sigma^{(n)}$ of E
brighter than $M_R=-21.5$ is significantly lower than the $\sigma^{(n)}$
of fainter E (0.99 confidence level, according to an F-test).
On the other hand, this magnitude dependence is not
found for the galaxies of other morphological classes. 
 
\begin{figure*} 
\vbox 
{\psfig{file=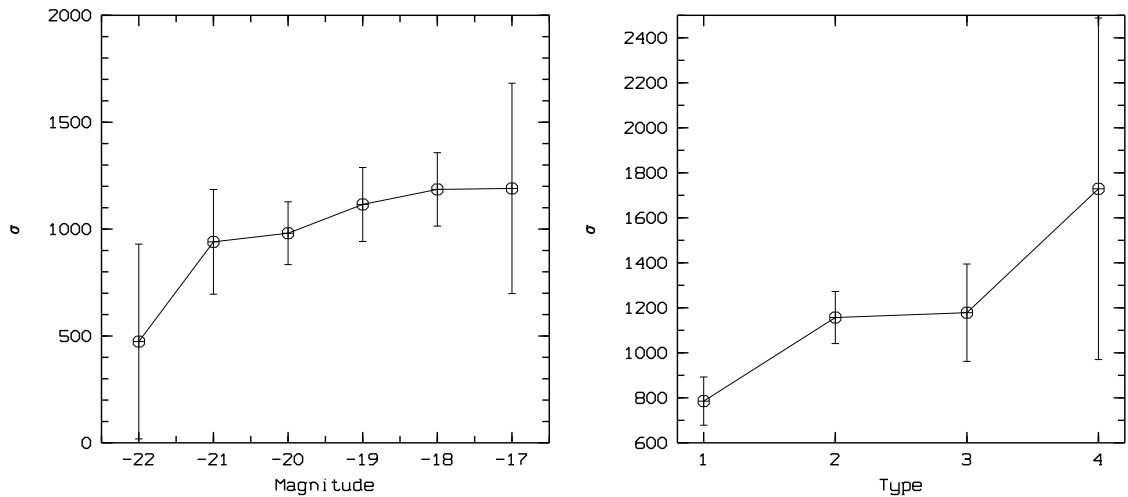,width=16.0cm,angle=270}} 
\caption[]{Segregation in the Coma cluster; left 
panel: $\sigma$ (km/s) vs. $M_R$ magnitude; right panel: 
$\sigma$ (km/s) vs. morphological type, with 1: E, 2: SO, 3: Se and 4: Sl.} 
\label{} 
\end{figure*} 
 
We then computed $\sigma^{(n)}$ separately for E, S0, Se and Sl 
of all luminosities (see Fig.~1b).
In general, later morphological type galaxies have a larger 
velocity dispersion than early type galaxies, 
although there is no significant difference between the S0 
and the Se $\sigma$. The E $\sigma$ is 15~\%
lower than the S0$+$Se $\sigma$, and the latter is 
20~\% lower than the Sl $\sigma$. An F-test qualifies these
differences as significant at more than 0.999 confidence level.
 
We have then considered the Coma (A~1656) cluster independently.
In order to achieve a higher signal-to-noise, we used in this case
bins of 1~mag, instead of 0.5~mag. We do not need to consider normalized
velocities in this case, as all galaxies belong to the 
same cluster.

Similarly to what was 
found for the synthetic cluster sample, in the Coma cluster
there is evidence for both luminosity and morphology-based segregation in 
velocity (see Fig.~3). In this case however, the
velocity dispersion seems to monotonically increase with magnitude
up to the fainter magnitude bin, where $\sigma \simeq 1200$~km~s$^{-1}$.
The difference between the $\sigma$ of galaxies in the brightest magnitude bin 
and that of galaxies in the adjacent magnitude bin, is consistent with the 
energy equipartition model, but the global trend is not.


As far as the morphological segregation is concerned, in general
later types seem to have larger $\sigma$, but the only significant
difference is between the $\sigma$ of E and the $\sigma$ of other galaxies.
  
\subsection{Spatial segregation}

\begin{figure} 
\vbox 
{\psfig{file=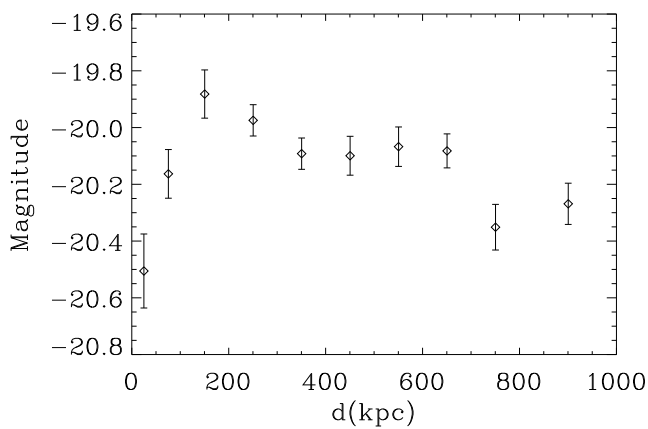,width=9.0cm}} 
\caption[]{The average absolute magnitude vs. 
clustercentric distance (in h$^{-1}$~kpc) for the synthetic cluster}
\label{} 
\end{figure} 

In the analysis of the spatial segregation, we have to take into account 
the fact that our sample is not magnitude complete. Magnitude incompleteness is
not a problem for the analysis of segregation in velocity, since velocities
and magnitudes of cluster galaxies are mostly uncorrelated. On the other hand,
observational biases may induce a spurious variation of the average galaxy 
magnitude with clustercentric distance that may be taken for evidence of
spatial luminosity segregation. In particular: (i) the limiting
magnitude of galaxies selected for observations with multi-fiber spectroscopy 
is generally brighter in fields of higher galaxy density (i.e. in the cluster
centres); (ii) deep observations generally only cover small fields of view,
i.e. the cluster centres and not the wider external regions. Note however that
the two biases result in opposite selection effects.

\begin{figure*} 
\vbox 
{\psfig{file=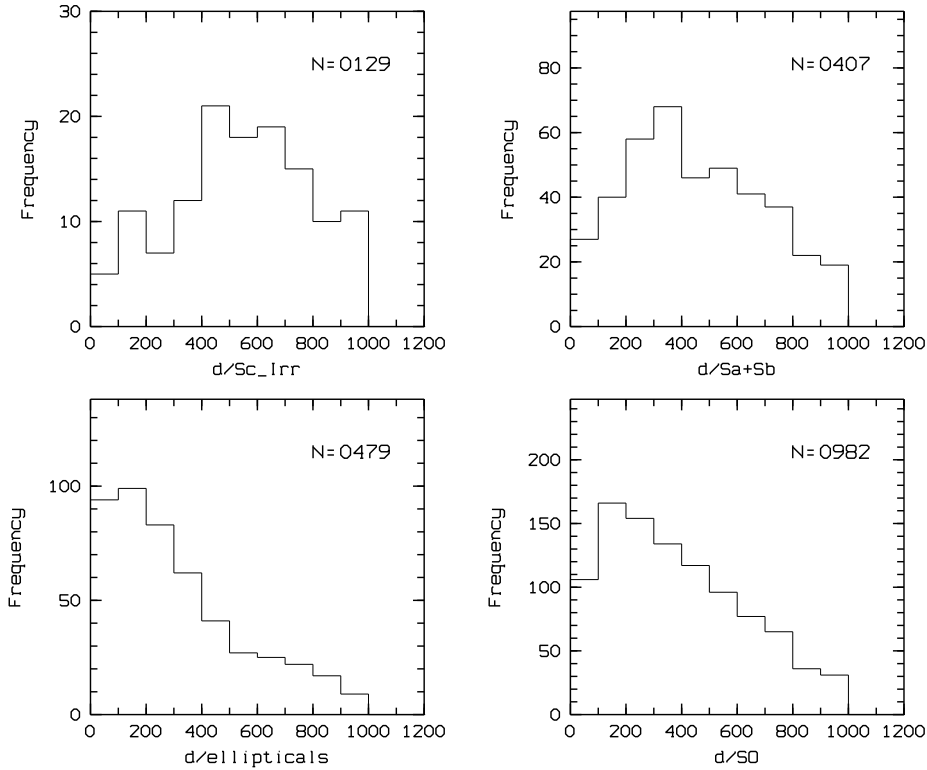,width=16.0cm,angle=270}} 
\caption[]{Histograms of the distance from the center in h$^{-1}$~kpc for 
the four synthetic cluster morphological sub-classes; bottom-left panel:
elliptical galaxies, bottom-right panel: S0 galaxies, top-right panel:
early spiral galaxies, top-left panel: late spiral galaxies. The number of
galaxies in each class is labelled.}
\label{} 
\end{figure*} 
 
We think that these biases do not seriously affect our analysis for the
following reasons:
(i) the ENACS sample (which constitutes a large part of our
sample) is free of these observational biases (Katgert et al. 1997);
(ii) our analysis is limited to the inner cluster regions, where the sampling
may be less inhomogeneous; (iii) the combination of several samples is likely 
to further reduce the problem, unless all samples are biased the same way.

In Fig.~4 we plot the average magnitude vs. clustercentric distance. Note that
the total number of galaxies per bin ranges from 92 to 316, large enough to
provide an adequate sampling of the luminosity function at the bright end. 
Galaxies in the central bin are clearly
brighter than those in the adjacent bin. In other bins the magnitude
decreases towards the outer parts of the (synthetic) cluster. 
It is difficult to see how observational biases can produce an artificial
decrease of the magnitude with increasing clustercentric distance, 
while leaving, at the
same time, an opposite trend just in the very centre of a cluster. It is
possible that a real effect, spatial segregation of the very bright galaxies,
is seen superposed on a more general trend that is created by an
observational bias. The evidence for spatial segregation must however be
considered as tentative only.

\begin{table} 
\caption[]{Biweight mean clustercentric distances and core-radii
$r_c$, in h$^{-1}$~kpc, of the fitted $\beta$ models for elliptical (E),  
lenticular (S0), early spiral (Se) and late spiral galaxies (Sl)} 
\begin{flushleft} 
\small 
\begin{tabular}{ccc} 
\hline 
\noalign{\smallskip} 
Type & $\overline{d}$ & $r_c$ \\ 
\hline 
\noalign{\smallskip} 
E  & 276$\pm$11 &  56$\pm$142 \\ 
S0 & 373$\pm$8  & 132$\pm$111 \\ 
Se & 451$\pm$13 & 159$\pm$243 \\ 
Sl & 549$\pm$23 & 197$\pm$174 \\ 
\noalign{\smallskip} 
\hline	    
\normalsize 
\end{tabular} 
\end{flushleft} 
\label{t-data3} 
\end{table} 
 
For what concerns the morphological segregation, 
we have considered the distributions of clustercentric distances of E, S0, 
Se and Sl, separately (see Fig.~5).
A Kruskall-Wallis test (see, e.g., Ledermann 1982) indicate that
the four distributions are significantly different ($>$0.999 confidence level). 
In Table~2 we list the average clustercentric distances of galaxies in the four 
morphological classes, and the core-radii of the King (1962) functions that best
fit the four distributions. Both the average clustercentric distance and the
core-radius increase along the Hubble sequence (E-S0-Se-Sl).

Note that the evidence for morphological segregation cannot be accounted for
by the trend of galaxy magnitudes with clustercentric distances, since
the average magnitudes of E, S0, Se and Sl are similar ($-20.2, -20.0, 
-20.2, -20.0$, respectively).
 
\begin{figure} 
\vbox 
{\psfig{file=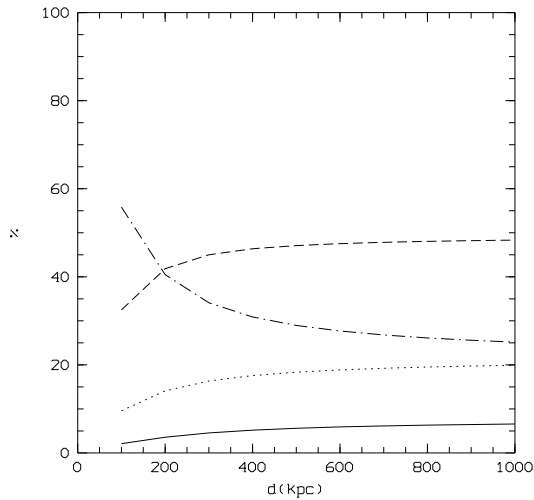,width=8.0cm,angle=270}} 
\caption[]{Percentage fraction 
of all galaxies in the synthetic cluster per morphological
type, within spheres of radii d, in h$^{-1}$~kpc.
Solid line: Sl; dotted line: Se; dashed line: S0; dash-dotted line: E.}
\label{} 
\end{figure} 

 
\begin{figure} 
\vbox 
{\psfig{file=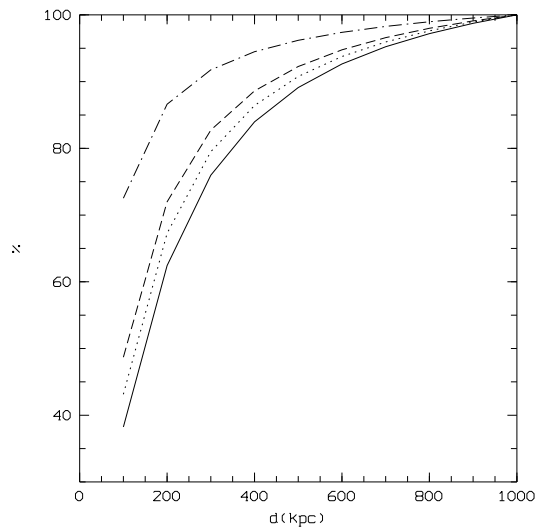,width=8.0cm,angle=270}} 
\caption[]{Cumulative fraction (in percent) of all galaxies of a given
morphological type within spheres of radii d, in h$^{-1}$~kpc.
Solid line: Sl; dotted line: Se; dashed line: S0; dash-dotted line: E.}
\label{titi} 
\end{figure} 
 
Via the Abel inversion, we have deprojected the King 
functions to determine the
relative fractions of the four morphological types within spheres of given 
radius; these are plotted in Fig.~6. In Fig.~7 we plot the integrated 
number of galaxies in each class within a sphere of given radius, normalised
to the total number of galaxies in that same class.
In addition to the well known morphological segregation between E, S0 and
spirals (see, e.g., Whitmore 1994), we also show here the segregation between
Se and Sl.


We found similar results by considering the Coma cluster sample alone.
 
\subsection{Phase-space segregation}

In the previous sections we have shown that galaxies of different morphological
types have different $\sigma$ (\S~3.1) and different spatial distributions
(\S~3.2). The size of our data-sample allowed us to look for morphological 
segregation in phase-space, through the determination of the \vdp of each
different morphological class. Each \vdp was determined using the
{\em LOWESS} technique (Gebhardt et al. 1994; see B97 for a recent 
application of this technique to galaxy clusters). 

In Fig.~8 we plot the \vdp 
of all the galaxies of the synthethic cluster, and the \vdp of E, S0, Se and Sl,
separately. Confidence levels are not shown, for clarity, 
but they are shown in Fig.~9, where we plot the \vdp for each class
(with models superposed, see below). The all-galaxy \vdp is similar
to the E-\vdp and the S0-\vdp, not surprisingly so, since most cluster galaxies
are E and S0. There are notable differences among the \vdp of the different 
classes:
\begin{itemize}
\item the late-type galaxies have a decreasing \vdp in the
centre, while the early-type galaxies have an increasing \vdp out to
0.2~h$^{-1}$~Mpc;
\item the \vdp of E, S0 and Se are similar for distances $\geq
0.2$~h$^{-1}$~Mpc, and nearly flat, while the Sl \vdp is different and
decreasing.
\end{itemize}

\begin{figure*} 
\vbox 
{\psfig{file=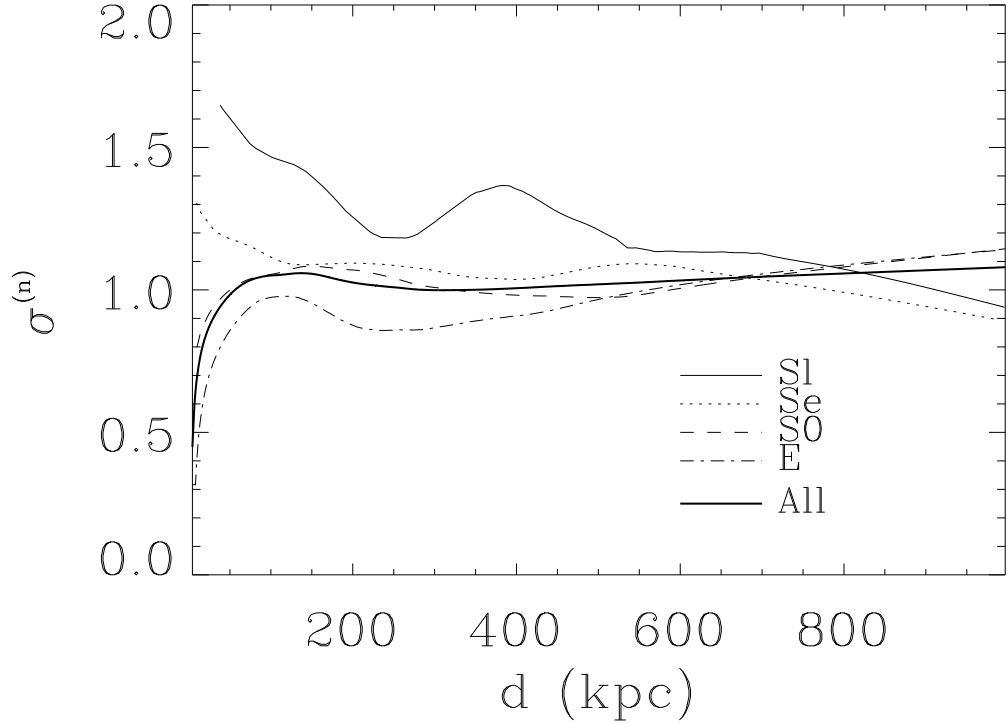,width=16.0cm}} 
\caption[]{The velocity dispersion profile of all galaxies (1997) of the 
synthetic cluster (heavy line), of E only (dash-dotted line), of S0 only 
(dashed line), of Se only (dotted line) and of Sl only (solid line).}
\label{} 
\end{figure*} 

\begin{figure*} 
\vbox 
{\psfig{file=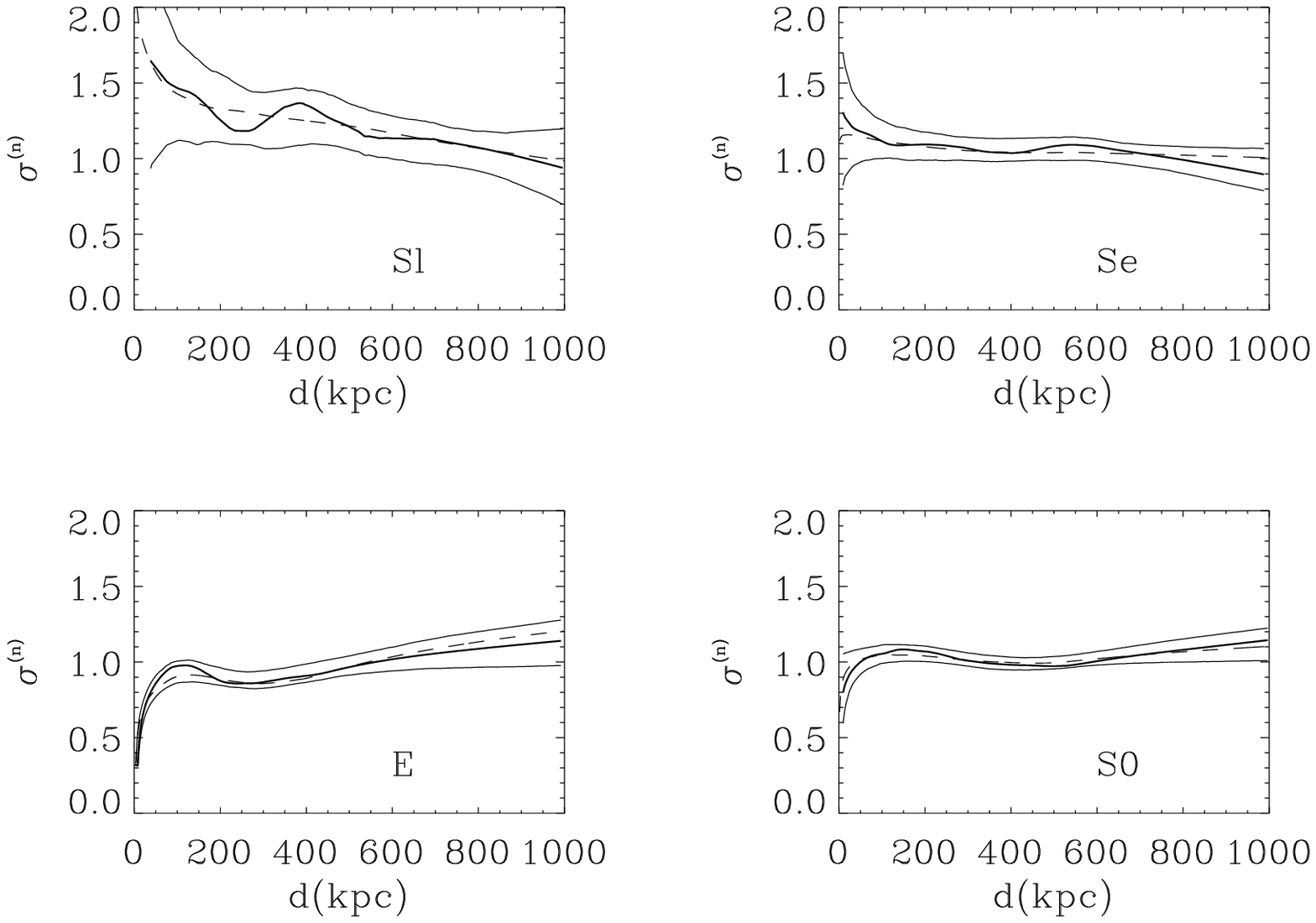,width=16.0cm}} 
\caption[]{The velocity dispersion profile of galaxies of a given morphological
class (heavy lines) 
with confidence levels at 90~\% (solid lines) and the
corresponding best fit models superposed (dashed lines; see text for further
details on the models); bottom-left panel: E (479 galaxies); bottom-right 
panel: S0 (982 galaxies); top-right panel: Se (407 galaxies); top-left panel: 
Sl (129 galaxies).}
\label{} 
\end{figure*} 

Interpreting these differences requires some modelling of the \vdp. We have
fitted the \vdp using simple kinematical models in which we take into account
the 3D galaxy distribution, as obtained from the King 
profiles (see \S~3.2),
and in which we allow the radial velocity profile to vary linearly with 
clustercentric distance, while the anisotropy, 
${\cal A} \equiv 1-(\sigma_t/\sigma_r)^2$, is allowed to vary only when
a constant anisotropy model could not provide a reasonable fit.


We found that the steep rise of the E \vdp requires a substantial degree of
tangential anisotropy in the centre (${\cal A} \simeq -7$)
and more or less isotropic orbits
outside 0.2--0.3~h$^{-1}$~Mpc. The same is true 
for the S0 \vdp, although in this case only a modest tangential anisotropy 
is needed in the cluster centre (${\cal A} \simeq -1$).
On the other hand, both the Se and the Sl \vdp are sufficiently well fitted
by constant anisotropy models (with decreasing radial velocity dispersion with
radius), with ${\cal A} \simeq 0.2$ for Se, and ${\cal A} \simeq 0.6$ for Sl. 

 
\section{Discussion} 
Based on a data-set of $\sim 2000$ galaxies in 40 Abell clusters,
we have found that the galaxy phase-space distributions depend on the
galaxy morphological types and luminosities. These results are based on a
synthetic cluster, generated by
combining all available cluster samples, and also on
the Coma cluster sample alone, for which enough data are available to allow
such an analysis. 

Both the galaxy velocity dispersion
and the average clustercentric distance are smaller, on average, for earlier 
type galaxies; galaxies brighter than $M_R=-21.5$ have a smaller velocity
dispersion than the other cluster galaxies, and the average magnitude of 
galaxies within 0.1~h$^{-1}$~Mpc from the cluster centre is $\sim 0.5$~mag 
lower than the average magnitude of the other cluster galaxies (although the
statistical significance of this last result is not well established, as it may
be biased by the incompleteness of our sample). 
Note that the decrease of $\sigma$ with magnitude only concerns E.

These findings confirm previous ones: the evidence for morphological
segregation dates back to Curtis (1918) and Humason et al. (1956), 
with regards to the spatial distribution, and to
Tammann (1972), Melnick \& Sargent (1977), and Moss \& Dickens (1977) with
regards to the velocity distribution. The evidence for luminosity
segregation dates back to Rood \& Turnrose (1968), with regards to the
spatial distribution, and to Rood et al. (1972) with regards to the velocity
distribution (see \S~1 for reference to more recent works).

The large size of our data-set has allowed us to go beyond these early results
and get a better insight into these topics:
\begin{enumerate}
\item Within the spiral class, we have considered separately 
early-spirals and late-spirals $+$ irregulars, and have shown that their 
distributions are different.
\item We have combined the spatial and velocity distributions to determine the
\vdp of each morphological class, and have shown that they are different.
\end{enumerate}

Modelling of these \vdp has allowed us to show that the E and S0 orbits are
predominantly tangential in the cluster core, and isotropic outside, while 
Se and Sl orbits have some degree of radial anisotropy (more marked for Sl).
The central 3D velocity dispersions
(in normalized units) are 1.4, 1.8, 2.1 and 2.4 for E, S0, Se and Sl
respectively, and the Sl 3D-$\sigma$ is larger than the E, S0 and Se
3D-$\sigma$ at all radii. The Se and Sl \vdp are steeper than the E and S0
\vdp; a peaked density profile can produce a steep \vdp
(den Hartog \& Katgert 1996), but in our sample we see the opposite, late-type
galaxies are characterized by {\em flatter} density profiles and at the
same time {\em steeper} \vdp than early-type galaxies. More likely, the
different morphological-type populations do not share the same dynamical status:
if S0 are the virialized cluster population, the
fact that the Sl kinetic energy is almost twice the S0 kinetic energy in the
cluster centre indicates that Sl are a bound yet not virialized cluster 
population. Their radial anisotropy then suggests 
that Sl may be falling into the cluster for the first time, similarly to
what has been suggested for narrow-angle-tail radio sources (O'Dea et al. 
1987). These conclusions recall those of B97 about
emission-line galaxies, and those by Tully \& Shaya (1984) about spirals in 
the Virgo cluster.

On the other hand, the low velocity dispersion of E,
their tangential anisotropy, and their central location,
are all suggestive of a phase-space distribution modified by the process of
dynamical friction (den Hartog \& Katgert 1996 and references therein).
Indeed, if the mass-to-light ratio of cluster galaxies is not a steep
function of their luminosities, it is easy to show that the timescale
of the dynamical friction process is $\simeq 0.2$~Gyr for 
$M_R \leq -21.5$ ellipticals in the cluster core (Sarazin 1986).
Consistently with this scenario, faint E ($M_R \geq -20$), i.e. those
less affected by the process of dynamical friction (assuming they are also less
massive), do not share the same \vdp as bright E, their orbits being more
isotropic and their \vdp more isothermal (a different kinematics for bright
and faint galaxy in the Coma cluster was already suggested by
Biviano et al. 1996).

\section{Conclusions}
Our analysis of the phase-space distributions of $\sim 2000$
cluster galaxies has shown
that they differ according to the galaxy luminosities and morphological 
types. We have interpreted these differences as evidence that 
the elliptical phase-space
distribution is evolving towards energy equipartition (e.g. Capelato et
al. 1981), while, at the other extreme, the late spirals are still falling
into the cluster for the first time
(e.g. Melnick \& Sargent 1977, Moss \& Dickens
1977, B97). The phase-space distributions of S0 and early
spirals are intermediate between these two extremes. 

These findings constrain theories of cluster formation and evolution, as they
suggest an ongoing evolution of the cluster population through the process of
dynamical friction and of secondary infall. The comparison of the phase-space
distributions of distant cluster galaxies with local ones is needed to shed
light on this topic. Observations with the {\em Hubble Space Telescope} 
(e.g. Dressler \& Smail 1997) and/or accurate spectroscopy on 8~m class 
telescopes should make such a comparison feasible soon.

\begin{acknowledgements}

{We wish to thank Dr.~Leo Metcalfe for a careful reading of the manuscript. 
We acknowledge financial support from the French GDR Cosmologie and INSU. 
AB acknowledges the hospitality of the Leiden Sterrewacht,
during his postdoctoral
fellowship, when the catalogue used in this paper was compiled.}

\end{acknowledgements}

\vfill 
\end{document}